\documentclass[rspublic]{article}
\usepackage[round]{natbib}
\usepackage{epsfig}
\usepackage{subfigure}
\usepackage{amssymb}
\usepackage{amsmath}
\usepackage{color}
\newtheorem{rem}{Remark}[section]
\newcommand{\br}{\begin{rem}}
\newcommand{\er}{\end{rem}}
\newtheorem{ex}[rem]{Example}
\newcommand{\bex}{\begin{ex}}
\newcommand{\eex}{\end{ex}}
\newtheorem{Def}[rem]{Definition}
\newcommand{\bd}{\begin{Def}}
\newcommand{\ed}{\end{Def}}
\newtheorem{theorem}[rem]{Theorem}
\newcommand{\bt}{\begin{theorem}}
\newcommand{\et}{\end{theorem}}
\newtheorem{lemma}[rem]{Lemma}
\newcommand{\bl}{\begin{lemma}}
\newcommand{\el}{\end{lemma}}
\newcommand{\be}{\begin{equation}}
\newcommand{\ee}{\end{equation}}
\newcommand{\bea}{\begin{eqnarray}}
\newcommand{\eea}{\end{eqnarray}}

\newcommand{\nn}{\nonumber}
\newcommand{\adots}{\mathinner{\mkern2mu\raise1pt\hbox{.}\mkern2mu
\raise4pt\hbox{.}\mkern2mu\raise7pt\hbox{.}\mkern1mu}}

\title{Mutation-Periodic Quivers, Integrable Maps and
             Associated Poisson Algebras\thanks{This article is dedicated to
             the memory of Robin Bullough.}}
\author{Allan P. Fordy ,  \\ School of Mathematics, \\
University of Leeds. \\ Leeds LS2 9JT, UK.\\
e-mail: amt6apf@leeds.ac.uk}

\date{23rd October, 2010}

\begin{document}

\maketitle

\bibliographystyle{plainnat}

\begin{abstract}
We consider a class of map, recently derived in the context of cluster
mutation.  In this paper we start with a brief review of the quiver context,
but then move onto a discussion of a related Poisson bracket, along with the
Poisson algebra of a special family of functions associated with these maps.  A
bi-Hamiltonian structure is derived and used to construct a sequence of Poisson
commuting functions and hence show complete integrability. Canonical
coordinates are derived, with the map now being a canonical transformation with
a sequence of commuting invariant functions. Compatibility of a pair of these
functions gives rise to Liouville's equation and the map plays the role of a
B\"acklund transformation.
\end{abstract}

\section{Introduction}

Robin Bullough's famous diagram represents a vast array of areas in Mathematics
and Mathematical Physics, together with a ``neural network'' of connections
(solid lines when established, dotted when expected) between them.  This
``Grand Synthesis of Soliton Theory'' shows that some remarkable connections
between seemingly disparate subjects have come about through the developments
of Integrable Systems, which have taken place in the last 40 years or so. Of
course, the diagram perpetually evolves, as dotted connections become solid and
as new subject areas (with corresponding links) are added to the array. In this
paper, I present some connections with subjects that didn't even exist until
recently!

Specifically, the present paper is concerned with {\em integrable maps} which
arise in the context of {\em cluster mutations} (see \cite{02-3}).  This gives
a connection to maps with the Laurent property, with the archetypical example
being the Somos $4$ iteration (see ``The On-Line Encyclopedia of Integer
Sequences'' at \cite{09-1}), which arises in the context of elliptic
divisibility sequences in number theory. In \cite{f09-1} we considered a class
of quiver which had a certain periodicity property under ``quiver mutation''.
The corresponding ``cluster exchange relations'' then give rise to sequences
with the Laurent property, which generalise many of the well known examples.

In this paper I first explain some of this background, but the main emphasis
will be on some associated Poisson algebras (with respect to an invariant
Poisson bracket of log-canonical type).  In terms of {\em canonical} variables
we obtain Hamiltonians (invariant under the action of the map) in exponential
form. The compatibility of one particular pair of invariant Hamiltonians leads
to Liouville's equation for each $q_i$, with the map now playing the role of a
B\"acklund transformation, bringing us back to one of the original ideas in
Soliton Theory!

\section{The Laurent Property}

The Somos $4$ sequence is generated by the iteration on the real line
\be\label{s4-sequ}  %
x_n x_{n+4} = x_{n+1}x_{n+3}+x_{n+2}^2,\quad\mbox{with}\quad x_0=x_1=x_2=x_3=1,
\ee  %
giving
$$
1,1,1,1,2,3,7,23,59,314,1529,8209,\dots
$$
Since we must divide by $x_n$ at each step, it is not obvious that we generate
integers from $x_8=59$ onwards. Even more, starting with initial conditions
$x_0=s, x_1=t, x_2=u, x_3=v$, we find that each term $x_n$ is a {\em Laurent
polynomial} in these initial values (ie, a polynomial in $s^{\pm 1},\, t^{\pm
1},\, u^{\pm 1},\, v^{\pm 1}$).  Integrality of the above numerical sequence
then follows by setting $r=s=u=v=1$. Considering an obvious generalisation of
(\ref{s4-sequ}) (called the Somos $N$ sequence)
$$
x_nx_{n+N}=\sum_{r=1}^{[N/2]} x_{n+r}x_{n+N-r},
$$
it is found that it too has the Laurent property for $N=4,5,6,7$, but fails at
$N=8$. {\em Failure} is rather simple to prove, since non-integer (rational)
elements occur fairly soon in the sequence.  Ad-hoc proofs exist for various
sequences (see \cite{91-17} for a proof in the case of Somos $4$), and can
often be adapted to other sequences. However, a remarkable (but complicated!)
proof for a very broad class of iteration was given in \cite{02-2}.

At about the same time, {\em cluster algebras} were introduced in \cite{02-3},
and it was shown that any map which arose as a {\em cluster exchange relation}
necessarily had the Laurent property.  Cluster algebras are an abstraction of
structures which arise in the study of total positivity of matrices and in the
canonical basis of a quantum group.  However, for this paper we need none of
this background. Neither do we need the full definition of a cluster algebra.
The most important aspect for us is the association with {\em quivers} and {\em
quiver mutation}.

\subsection{Quiver Mutation}

A quiver is a {\em directed graph}, consisting of $N$ nodes with directed edges
between them.  There may be several arrows between a given pair of vertices,
but for cluster algebras there should be no $1$-cycles (an arrow which starts
and ends at the same node) or $2$-cycles (an arrow from node $a$ to node $b$,
followed by one from node $b$ to node $a$). A quiver $Q$, with $N$ nodes, can
be identified with the unique skew-symmetric $N\times N$ matrix $B_Q$ with
$(B_{Q})_{ij}$ given by the number of arrows from $i$ to $j$ minus the number
of arrows from $j$ to $i$.

\begin{figure}[hbt]
\begin{minipage}[c]{5cm}
\centering
\includegraphics[width=3cm]{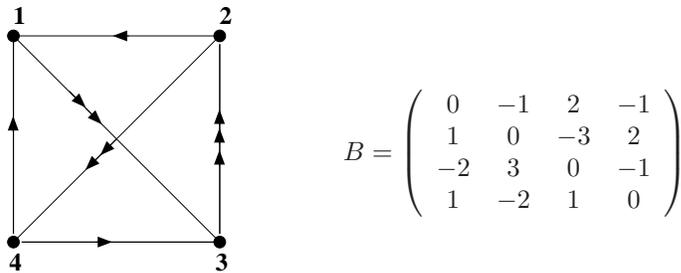}
\end{minipage}
\quad
\begin{minipage}[c]{4cm}
$$
B= \left(
\begin{array}{cccc}
  0 & -1 & 2 & -1 \\
 1 & 0 & -3 & 2 \\
  -2 & 3 & 0 & -1 \\
  1 & -2 & 1 & 0 \\
\end{array}\right)
$$
\end{minipage}
\caption{The Somos $4$ quiver $S_4$ and its matrix.}\label{s4quivermatrix}
\end{figure}

\begin{figure}[hbt]
\begin{minipage}[c]{5cm}
\centering
\includegraphics[width=3cm]{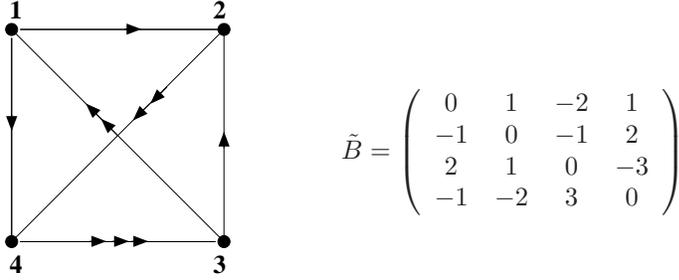}
\end{minipage}
\quad
\begin{minipage}[c]{4cm}
$$
\tilde B= \left(
\begin{array}{cccc}
  0 & 1 & -2 & 1 \\
 -1 & 0 & -1 & 2 \\
  2 & 1 & 0 & -3 \\
  -1 & -2 & 3 & 0 \\
\end{array}\right)
$$
\end{minipage}
\caption{Mutation $\tilde S_4=\mu_1 S_4$ of the quiver $S_4$ at node $1$ and
its matrix.}\label{s4quiverbmatrix}
\end{figure}

An important quiver for our discussion is the one corresponding to the Somos
$4$ sequence (both quiver and matrix in Figure \ref{s4quivermatrix}).
\bd[Quiver Mutation]\label{d:mutate}   %
Given a quiver $Q$ we can mutate at any of its nodes. The mutation of $Q$ at
node $k$, denoted by $\mu_k Q$, is constructed (from $Q$) as follows:
\begin{enumerate}
\item  Reverse all arrows which either originate or terminate at node $k$.
\item  Suppose that there are $p$ arrows from node $i$ to node $k$ and $q$
arrows from node $k$ to node $j$ (in $Q$). Add $pq$ arrows going from node $i$
to node $j$ to any arrows already there. \item Remove (both arrows of) any
two-cycles created in the previous steps.
\end{enumerate}
\ed    %
Note that in Step $2$, $pq$ is just the number of paths of length $2$ between
nodes $i$ and $j$ which pass through node $k$.
\br[Matrix Mutation]   %
Let $B$ and $\tilde B$ be the skew-symmetric matrices corresponding to the
quivers $Q$ and $\tilde Q=\mu_k Q$.  Let $b_{ij}$ and $\tilde b_{ij}$ be the
corresponding matrix entries.  Then quiver mutation amounts to the following
formula
\be   \label{gen-mut}  %
\tilde b_{ij}= \left\{ \begin{array}{ll}
                        -b_{ij} & \mbox{if}\;\; i=k\;\;\mbox{or}\;\; j=k, \\
                        b_{ij}+\frac{1}{2} (|b_{ik}|b_{kj}+b_{ik}|b_{kj}|)
                        & otherwise.
                        \end{array}  \right.
\ee    %
\er      %
It is an exercise to show that with these definitions, the Somos $4$ quiver and
matrix are transformed to those of Figure \ref{s4quiverbmatrix}, if we mutate
at node $1$.

\subsection{Cluster Exchange Relations}

Given a quiver (with $N$ nodes), we attach a variable at each node, labelled
$(x_1,\cdots ,x_N)$.  When we mutate the quiver we change the associated matrix
according to formula (\ref{gen-mut}) and, {\em in addition}, we transform the
cluster variables $(x_1,\cdots ,x_N)\mapsto (x_1,\cdots ,\tilde x_k,\cdots,x_N)$, where
\be  \label{ex-rel}   %
x_k \tilde x_k = \prod_{b_{ik}>0} x_i^{b_{ik}}+
            \prod_{b_{ik}<0} x_i^{-b_{ik}},
           \qquad   \tilde x_i = x_i \;\;\mbox{for}\;\; i\neq k .
\ee   %
If one of these products is empty (which occurs when all $b_{ik}$ have the
same sign) then it is replaced by the number $1$.  This formula is called the
(cluster) {\em exchange relation}.  Notice that it just depends upon the
$k^{th}$ column of the matrix.  Since the matrix is skew-symmetric, the
variable $x_k$ {\bf does not} occur on the right side of (\ref{ex-rel}).

After this process we have a new quiver $\tilde Q$, with a new matrix $\tilde
B$.  This new quiver has cluster variables $(\tilde x_1,\cdots ,\tilde x_N)$.
However, since the exchange relation (\ref{ex-rel}) acts as the identity on all
except one variable, we write these new cluster variables as $(x_1,\cdots
,\tilde x_k,\cdots ,x_N)$.  We can now repeat this process and mutate
$\tilde Q$ at node $\ell$ and produce a third quiver $\tilde{\tilde Q}$, with
cluster variables $(x_1,\cdots ,\tilde x_k,\cdots ,\tilde x_\ell,\cdots ,x_N)$,
with $\tilde x_\ell$ being given by an analogous formula (\ref{ex-rel}).

\br[Involutive Property of the Exchange Relation]   %
If $\ell=k$, then $\tilde{\tilde Q}=Q$, so we insist that $\ell\neq k$.
\er   %

\bex[The Somos $4$ Quiver $S_4$]  \label{s4exchange}  {\em  %
Placing $x_1,x_2,x_3,x_4$ respectively at nodes $1$ to $4$ of quiver $S_4$ (of
Figure \ref{s4quivermatrix}) gives the {\em initial cluster}.  Along with the
{\em quiver} mutation (leading to $\mu_1 S_4$ of Figure \ref{s4quiverbmatrix}),
we also have the {\em exchange relation}
\be\label{x1x1t}  %
x_1\tilde x_1 = x_2 x_4+x_3^2.
\ee  %
This corresponds to one arrow coming {\bf into} node $1$ from each of nodes $2$
and $4$ with $2$ arrows going {\bf out} to node $3$.

We can now consider mutations of quiver $\tilde S_4=\mu_1 S_4$.  To avoid too
many ``tildes'', let us write $\tilde x_1 = x_5$, so quiver $\tilde S_4$ has
$x_5,x_2,x_3,x_4$ respectively at nodes $1$ to $4$.  Mutation at node $1$ would
just take us back to quiver $S_4$ (as noted in the above remark).  We compare
the exchange relations we would obtain by mutating at nodes $2$ or $3$.

Mutation at node $2$ would lead to exchange relation
\be\label{x2x2t}  %
x_2\tilde x_2 = x_3 x_5+x_4^2,
\ee  %
whilst that at node $3$ would lead to
\be\label{x3x3t}  %
x_3\tilde x_3 = x_2 x_5^2+x_4^3.
\ee  %
We see that the right hand sides of formula (\ref{x1x1t}) and (\ref{x2x2t}) are
related by a {\bf shift}, whilst formula (\ref{x3x3t}) is entirely {\em
different}.  In fact, it can be seen in Figures \ref{s4quivermatrix} and
\ref{s4quiverbmatrix} that the configuration of arrows at node $2$ of quiver
$\tilde S_4$ is {\em exactly the same} as that at node $1$ of quiver $S_4$,
thus giving the {\em same} exchange relation.  In fact, we have more.  The {\em
whole quiver} $\tilde S_4$ is obtained from $S_4$ by just {\bf rotating the
arrows}, whilst keeping the nodes fixed.  It follows that mutation of quiver
$\tilde S_4$ {\em at node $2$} just leads to a further rotation, with node $3$
inheriting this same configuration of arrows.  If at each step we relabel
$\tilde x_n$ as $x_{n+4}$, the $n^{th}$ exchange relation can be written
\be\label{xnxnt}  %
x_n x_{n+4} = x_{n+1} x_{n+3}+x_{n+2}^2.
\ee  %
This rotational property of the quiver has lead to an {\bf iteration}, which in
this case is just Somos $4$.
}\eex  %

In \cite{f09-1} we introduced and studied quivers with this type of rotational
property. Consider the $N\times N$ matrix
$$
\rho =  \left( \begin{array}{cccc}
                   0 & \cdots & \cdots & 1 \\
                  1 & 0 & & \vdots \\
                     &\ddots & \ddots & \vdots \\
                      & & 1 & 0
                      \end{array} \right).
$$
The above rotation, which we write $S_4\rightarrow \tilde S_4=\rho S_4$, is
achieved in the matrix formulation by
$$
\tilde B = \rho B \rho^{-1},
$$
with $N=4$ in this case.

Consider a quiver $Q=Q(1)$, with $N$ nodes.  We consider a sequence of
mutations, starting at node $1$, followed by node $2$, and so on. Mutation at
node $1$ of a quiver $Q(1)$ will produce a second quiver $Q(2)$.  The mutation
at node $2$ will therefore be of quiver $Q(2)$, giving rise to quiver $Q(3)$
and so on. We define a {\em period $m$ quiver} as follows.
\bd   %
A quiver $Q$ has \textit{period $m$} if it satisfies $Q(m+1)=\rho^m Q(1)$ (with
$m$ the minimum such integer). The mutation sequence is depicted by
\be   \label{periodchain}   %
Q=Q(1) \stackrel{\mu_1}{\longrightarrow} Q(2) \stackrel{\mu_2}{\longrightarrow}
  \cdots \stackrel{\mu_{m-1}}{\longrightarrow} Q(m)
       \stackrel{\mu_m}{\longrightarrow} Q(m+1)=\rho^m Q(1),
\ee    %
and called the {\em periodic chain} associated to $Q$.

The corresponding matrices would then satisfy $B(m+1)=\rho^m B(1)\rho^{-m}$.
\ed  %

\br[The Sequence of Mutations]  %
We must perform the {\bf correct} sequence of mutations.  For instance, if we
mutate $\mu_1 S_4$ at node $3$, we obtain a quiver which has $5$ arrows from
node $4$ to node $1$, which {\bf cannot} be permutation equivalent to
$Q(1)=S_4$.  As we previously saw, the corresponding exchange relation
(\ref{x3x3t}) was also different.
\er  %

\br[Periodicity and Iterations]  %
{\bf Period $1$ quivers} correspond to iterations on the real line. {\bf Period
$m$ quivers} correspond to iterations on $\mathbb{R}^m$. The formula
(\ref{ex-rel}) consists of only two terms (additively), corresponding to {\bf
incoming} and {\bf outgoing} arrows.  Both the Somos $4$ and Somos $5$
iterations can be built in this way, but {\bf not} Somos $6$ or $7$, which
contain $3$ terms.
\er  %

In \cite{f09-1} we give a {\em full} classification of period $1$ quivers, a
partial classification of period $2$ quivers and examples of higher period
ones.

\subsection{Primitive Quivers}

\begin{figure}[htb]
\centering \subfigure[$P_4^{(1)}$]{
\includegraphics[width=2cm]{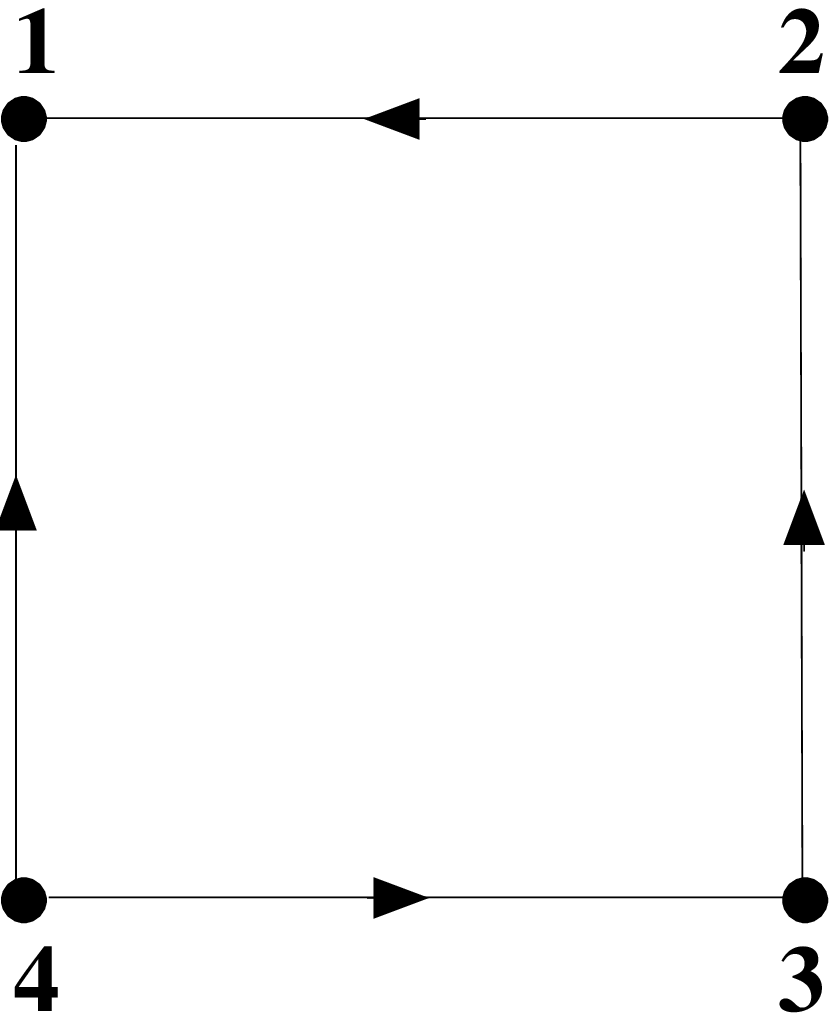}\label{subfig:P41}
}\qquad \subfigure[$P_4^{(2)}$]{
\includegraphics[width=2cm]{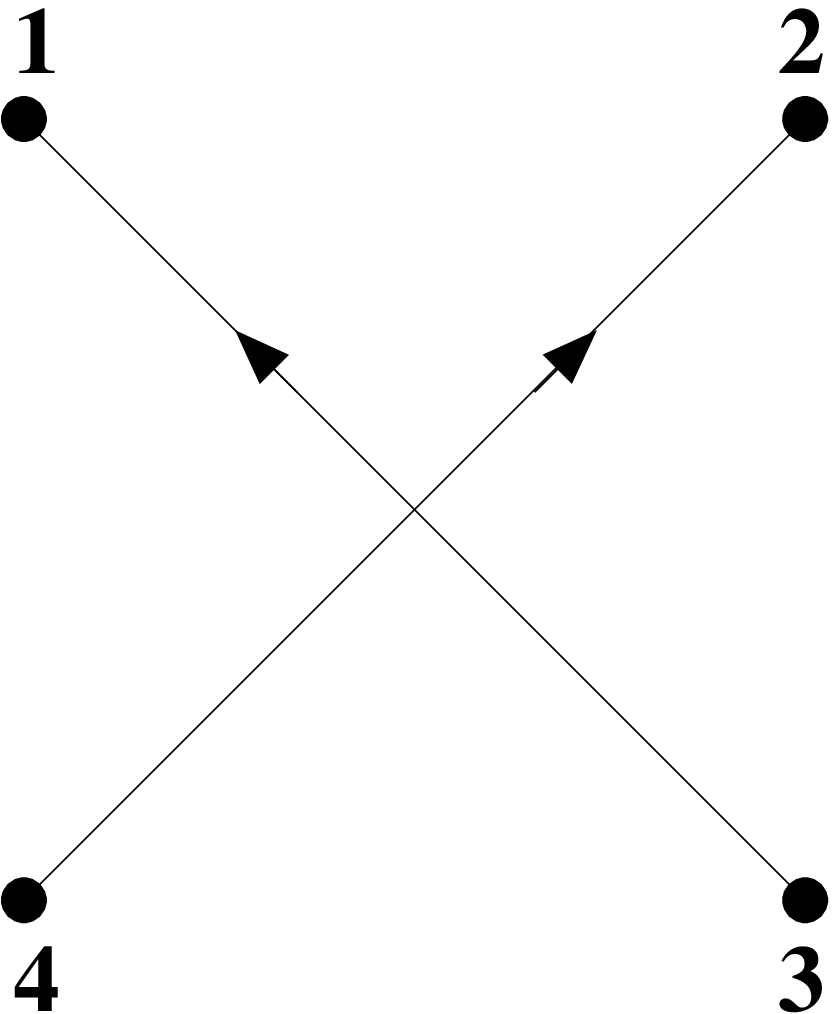}\label{subfig:P42}
}\qquad \subfigure[$P_5^{(1)}$]{
\includegraphics[width=2.5cm]{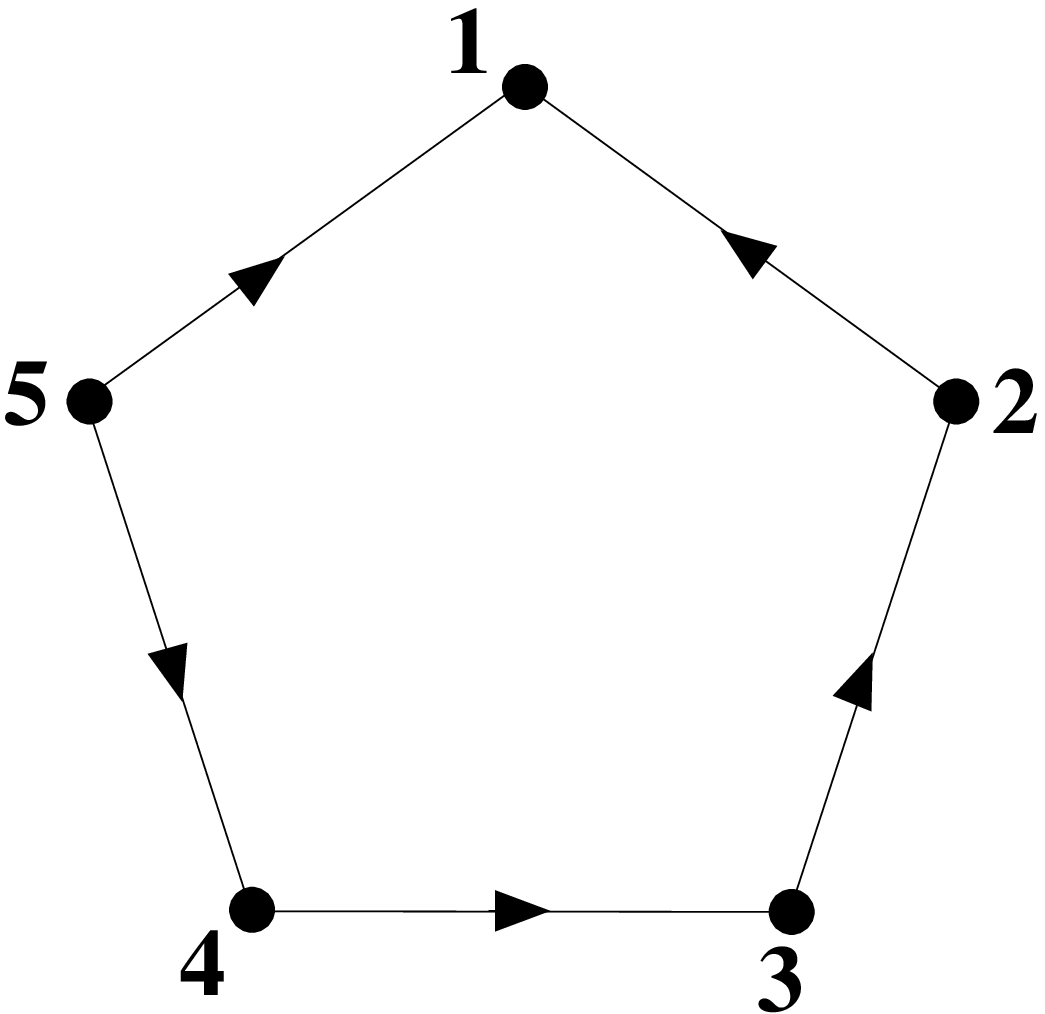}\label{subfig:P51}
}\qquad \subfigure[$P_5^{(2)}$]{
\includegraphics[width=2.5cm]{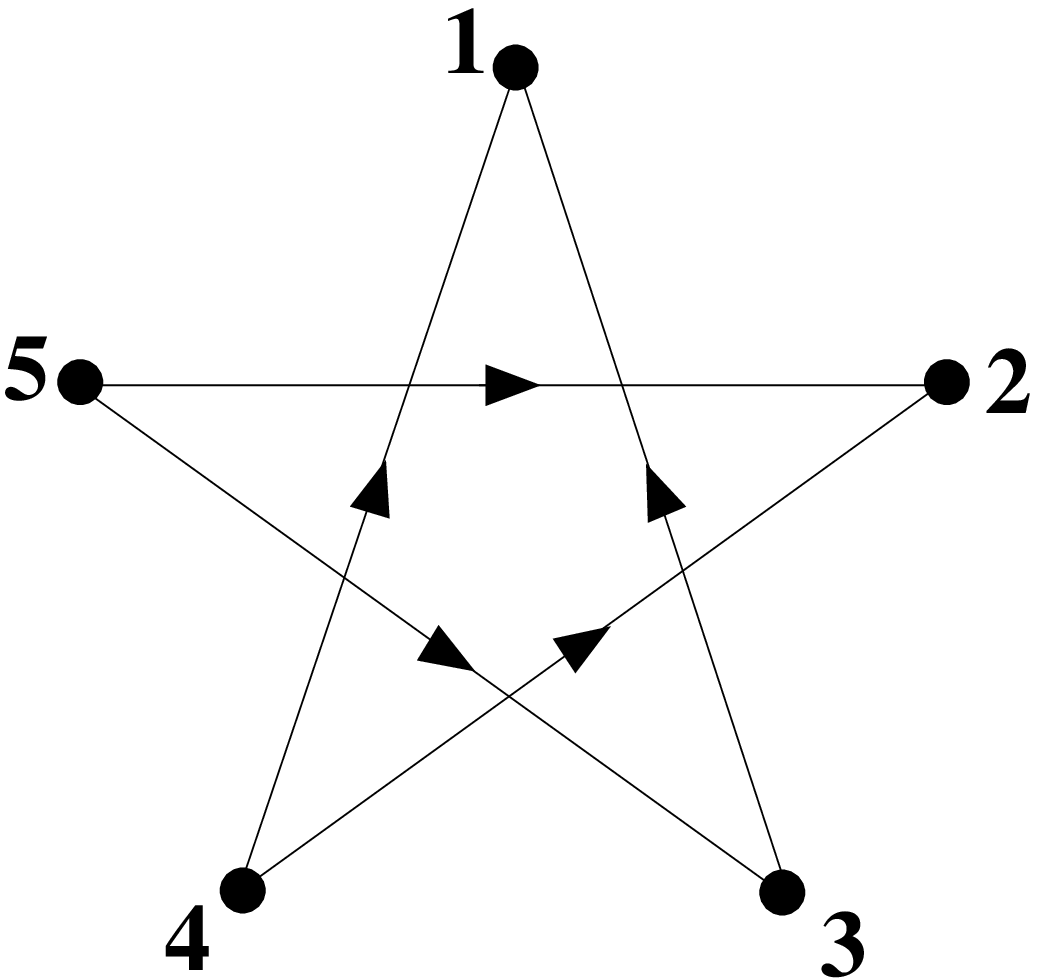}\label{subfig:P52}
} \caption{The period $1$ primitives for $4$ and $5$ nodes.} \label{45node}
\end{figure}

In our classification of period $1$ quivers we introduced a special class of
quivers, called {\em primitives}.  An important feature of a primitive is that
node $1$ is a {\em sink}, so only step $1$ of the mutation (Definition
\ref{d:mutate}) is needed.  We constructed a basis of primitives for each $N$
(see Figure \ref{45node} for $N=4,5$).
The basis consists of $P_N^{(r)}$, for $1\leq r\leq N/2$.

\begin{figure}[htb]
\centering \subfigure[$P_4^{(1)}$]{
\includegraphics[width=2cm]{P41.eps}
}\qquad \subfigure[$P_4^{(2)}$]{
\includegraphics[width=2cm]{P42.eps}
}\qquad \subfigure[$P_2^{(1)}$]{
\includegraphics[width=2cm]{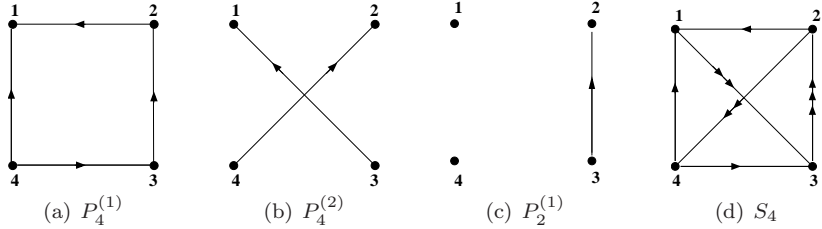}
}\qquad \subfigure[$S_4$]{
\includegraphics[width=2cm]{somos4quiver.eps}
}\caption{One of $P_4^{(1)}$ minus two of $P_4^{(2)}$ plus two of $P_2^{(1)}$
gives $S_4$}\label{s4sum}
\end{figure}

In our classification, the primitives are the ``atoms'' out of which we build
the general period $1$ quiver for each $N$.  For given $N$, we can start with
an arbitrary linear combination of $P_N^{(r)}$, with integer coefficients
$$
Q_N^0=\sum_{r=1}^{[N/2]} m_r P_N^{(r)}.
$$
If all the $m_r$ have the {\em same} sign, then $Q_N^0$ already has period $1$,
but otherwise, we must add ``correction terms'', which are integer combinations
of primitives with $N-2k$ nodes ($1\leq k\leq [N/2]$).  Our classification of
period $1$ quivers gives the formula for these coefficients in terms of the
original coefficients $m_r$.  For the Somos $4$ quiver, we have $m_1=1, m_2=-2$
and our formula requires the addition of a further two arrows between nodes $3$
and $2$ (see Figure \ref{s4sum}).

\subsection{Laurent Property vs Complete Integrability}

Each iteration we obtain through our construction is {\bf guaranteed} to have
the Laurent property (by the results of \cite{02-3}).  However, only special
cases are expected to be completely integrable in any sense (see \cite{91-4}
for various definitions).  For instance, the most general $4$ node, period $1$
quiver corresponds to the iteration
\be\label{gen4quiv}  %
x_n x_{n+4}=x_{n+1}^rx_{n+3}^r+x_{n+2}^s .
\ee  %
The iterations with $r=1, s\in\{0,1,2\}$ are analysed by \cite{07-2}, who shows
that these cases are Liouville integrable (even {\em super-integrable} when
$s=0,1$).

We first write (\ref{gen4quiv}) as a {\bf map} on the $4-$dimensional space
with coordinates $x_0,x_1,x_2,x_3$:
\be  \label{map4}   %
\varphi(x_0,\dots ,x_{3}) =
   \left(x_1,x_2 ,x_{3},\frac{x_1^r x_{3}^r+x_2^s}{x_0}\right).
\ee   %
The log-canonical Poisson bracket $\{x_i,x_j\}=P_{ij} x_ix_j$ (see \cite{03-5}
for a general discussion), where
\be\label{pb4}  %
                 P=\left(\begin{array}{cccc}
                 0 & r & s  & r(1+s) \\
                 -r  & 0 & r  & s \\
                 -s & -r  & 0 & r \\
                 -r(1+s) & -s & -r  & 0
                 \end{array}\right),
\ee  %
is {\em invariant} under the action of the map $\varphi$.  This means that, if
$\tilde {\bf x} = \varphi ({\bf x})$, then $\{\tilde x_i,\tilde x_j\} =
P_{ij}\tilde x_i \tilde x_j$.
\br  %
It is an interesting fact that the matrix $P$ is (up to an overall factor) the
{\bf inverse} of the $B$ matrix for the corresponding quiver. The factor is the
Pfaffian $(2+s)r^2-s^2$, which actually vanishes in the Somos $4$ case.
Nevertheless, the matrix $P$, with $r=1,s=2$, is still invariant under the map.
\er  %

Liouville integrability is defined in the same way as for continuous
Hamiltonian systems (see \cite{91-4}).  We must first use the Casimir functions
(when the Poisson matrix is degenerate) to reduce to the symplectic leaves,
whose dimension is $2d$, where $d$ is the number of degrees of freedom.  We
then require the existence of $d$, functionally independent Hamiltonians,
$h_1,\dots ,h_d$, which should be {\em in involution} (so $\{h_i,h_j\}=0$, for
all $i,j$).  For the discrete case, we have the extra requirement that the
functions $h_1,\dots ,h_d$ are {\em invariants} of the map.  This means that
the map has a system of $d$ commuting {\em continuous symmetries} (the
Hamiltonian flows).  In the continuous case we can say that the Hamiltonian
system is solvable, {\em up to quadrature}, but this notion is not carried to
the discrete case.

In \cite{07-2}, it is shown that for cases $r=1, s=0,1$, there are $3$
independent, invariant functions, out of which it is possible to construct two
Poisson commuting functions.  Such systems (with additional first integrals)
are known as {\em super-integrable}.  In the case $r=1,s=2$ (Somos $4$) the
Poisson bracket is degenerate, with two Casimir functions, which are {\em not
invariant} under the map.  However, the action of the map on these Casimirs is
by a $2-$dimensional integrable map (a special case of the symmetric QRT map
(see \cite{88-5})).

It is not known whether the map (\ref{gen4quiv}) is Liouville integrable for
other values of $r$ and $s$, but some of the standard integrability tests (such
as {\em algebraic entropy} (see \cite{99-12})) indicate {\bf
non-integrability}.
\br  %
Isolating and analysing the integrable cases is one of the most interesting
outstanding problems.
\er  %

\section{The $P_N^{(1)}$ Iteration as a Map}

The following iteration corresponds to the period $1$ primitive $P_N^{(1)}$
with $N$ nodes:
\be   \label{en}   %
x_n x_{n+N}=x_{n+1}x_{n+N-1}+1,
\ee  %
with initial conditions $x_i=a_i$ for $0\leq i\leq N-1$. In \cite{f09-1} it was
shown that there exists a special sequence of functions
\be  \label{jn}   %
J_{n}=\frac{x_n+x_{n+2}}{x_{n+1}}, \quad\mbox{satisfying}\quad J_{n+N-1}=J_{n}.
\ee  %
With the given initial conditions, we have $\{J_{i}=c_i:0\leq i\leq N-2\}$,
together with the periodicity condition, which can also be written as
$J_{n}=c_n$ with $c_{n+N-1}=c_n$.

\bt[Linearisation]  \label{linear-thm}  %
If the sequence $\{x_n\}$ is given by the iteration (\ref{en}), with initial
conditions $\{x_i=a_i: 0\leq i\leq N-1\}$, then it also satisfies
\be  \label{lineq}  %
x_n+x_{n+2(N-1)}=S_{N}\, x_{n+N-1},
\ee  %
where $S_{N}$ is a function of $c_0,\cdots ,c_{N-2}$, which is symmetric under
cyclic permutations.
\et  %

Here we restrict to the case of even $N$. The first few $S_{N}$ take the form
\bea   %
&& S_{2} = c_0;  \quad  S_{4} = c_0c_1c_2-c_0-c_1-c_2; \nn  \\
&& S_{6} = c_0c_1c_2c_3c_4-c_0c_1c_2-c_1c_2c_3-c_2c_3c_4-
               c_3c_4c_0-c_4c_0c_1 \nn\\
&&   \hspace{5cm} +c_0+c_1+c_2+c_3+c_4. \nn
\eea  %
The function $S_N$ is an {\em invariant function} of the nonlinear map
(\ref{en}), so this linearisation depends upon the particular initial
conditions.

\subsection{The Log-Canonical Poisson Bracket}\label{log-can-section}

As in the $4^{th}$ order case (\ref{map4}), write the $N^{th}$ order iteration
(\ref{en}) as a map of the space with coordinates $(x_0,\dots ,x_{N-1})$, given
by
\be  \label{mapN}   %
\varphi(x_0,\dots ,x_{N-1}) =
   \left(x_1,\dots ,x_{N-1},\frac{x_1 x_{N-1}+1}{x_0}\right).
\ee   %
Again we seek an invariant Poisson bracket of log-canonical form:
\be   \label{log-can}   %
\{x_i,x_j\} = P_{ij} x_i x_j, \quad  0\leq i<j\leq N-1 ,
\ee   %
for some constants $P_{ij}$.  We seek the value of these constants for which
this Poisson bracket is invariant under the action of the map $\varphi$.
Writing $\tilde {\bf x} = \varphi ({\bf x})$, we require
$$
\{\tilde x_i,\tilde x_j\} = P_{ij} \tilde x_i \tilde x_j.
$$
The shift structure of the map (\ref{mapN}) implies a banded structure, with
$P_{i+1\, j+1}=P_{ij}$, so the undetermined constants are $P_{0j},\, j=1,\dots
,N-1$.  The precise form of $\tilde x_{N-1}$ puts strong constraints on these,
which can be determined up to an overall multiplicative constant.

\bl \label{lemma-pij} %
For a nontrivial Poisson bracket of the form (\ref{log-can}) to be invariant
under the map (\ref{mapN}), we require $N$ to be {\bf even}, in which case the
coefficients take the form
$$
P_{ij}=\left\{ \begin{array}{ll}
    1 & \mbox{when} \;\; i<j\;\;\mbox{and}\;\; i+j  \;\;\mbox{is odd},  \\
    0 & \mbox{when} \;\; i<j\;\;\mbox{and}\;\; i+j  \;\;\mbox{is even}.
    \end{array}  \right.
$$
This Poisson bracket is non-degenerate.
\el  %

\br  %
Again, it is an interesting fact that the matrix $P$ is (up to an overall
factor) the {\bf inverse} of the $B$ matrix for the corresponding quiver.
\er  %

\subsection{The Poisson Algebra of Functions $J_m$}

The independent functions $J_0,\dots , J_{N-2}$, written in terms of the
coordinates $(x_0,\dots ,x_{N-1})$, are given by
\be  \label{jm}  %
J_m=\frac{x_m+x_{m+2}}{x_{m+1}},\;\;\; m=0,\dots ,N-3, \;\;\;
   J_{N-2} = \frac{x_0x_{N-2}+x_1x_{N-1}+1}{x_0x_{N-1}} .
\ee  %
Under the action of the map $\varphi$ they satisfy the cyclic conditions
\be  \label{jophi}  %
J_n\circ\varphi = J_{n+1},\;\; n=0,\dots ,N-3 \quad\mbox{and}\;\;
  J_{N-2}\circ\varphi = J_0.
\ee  %
We just need to calculate the $N-3$ brackets $\{J_0,J_n\},\; n=1,\dots ,N-3$,
since all others follow through the above relations.  These are easily
calculated to be
\be  \label{pbj0}  %
\begin{array}{l}
\{J_0,J_1\}= 2 J_0J_1-2, \quad
      \{J_0,J_{2m-1}\}= 2 J_0J_{2m-1}, \;\;\; 2\leq m\leq M-1,  \\[3mm]
 \{J_0,J_{2m}\}= -2 J_0J_{2m}, \;\;\; 1\leq m\leq M-2,
\end{array}
\ee   %
where $N=2M$.  The cyclic action of $\varphi$ then implies
\be  \label{pbjm}  %
\begin{array}{l}
\{J_m,J_{m+1}\}= 2 J_mJ_{m+1}-2, \;\;\; 1\leq m\leq N-3, \\[3mm]
\{J_m,J_n\}=2(-1)^{m+n-1}J_mJ_n
   \;\;\;\mbox{for}\;\;\; 1\leq m\leq n-2\leq N-4,  \\[3mm]
   \{J_0,J_{N-2}\}= -2 J_0J_1+2,
\end{array}
\ee   %
where the relation for $\{J_0,J_{N-2}\}$ was obtained from that of
$\{J_{N-3},J_{N-2}\}$ through the action of $\varphi$.

By taking cyclic sums of any function of the $J_n$ we can build functions which
are invariant under the action of $\varphi$.  Since the Poisson bracket
(\ref{log-can}) (with $P_{ij}$ as given in Lemma \ref{lemma-pij}) is
non-degenerate (on a $2M-$dimensional space), our task is to select $M$
invariant functions which are {\em in involution}.  It is, in fact, easier to
work with the Poisson bracket relations (\ref{pbj0}) and (\ref{pbjm}), which
define a Poisson bracket on the $(2M-1)$ dimensional $J-$space.  The
corresponding Poisson matrix $P$ is the sum of two homogeneous parts:
$P=P_2+P_0$, each of which is itself a Poisson matrix.  These therefore define
a compatible pair of Poisson brackets:
\bd[Compatible Poisson Brackets]\label{compat-pb} 
The matrices $P_0, P_2$ and $P=P_2+P_0$ define compatible Poisson brackets
$$
\{f,g\}_i = \nabla f P_i \nabla g , \quad i=0,2\quad\mbox{and}\quad
    \{f,g\}_P = \nabla f (P_2+P_0) \nabla g.
$$
\ed  %
We use these brackets to define a {\em bi-Hamiltonian ladder} (see
\cite{78-4}), starting with the Casimir function of $P_0$ and ending with that
of $P_2$:
\be   \label{biham}  %
(P_2+P_0)(\nabla h_M -\nabla h_{M-1} +\nabla h_{M-2}-\dots
         +(-1)^{M+1} \nabla h_1)=0,
\ee   %
where $h_k$ is a homogeneous polynomial of degree $2k-1$.  The homogeneity
property of $P_0, P_2, h_k$ leads to equation (\ref{biham}) decoupling into a
sequence of $M+1$ homogeneous equations (the bi-Hamiltonian ladder):
\be   \label{ladder}  %
P_0 \nabla h_1=0, \;\; P_0\nabla h_k = P_2 \nabla h_{k-1} ,
\;\;\;\mbox{for}\;\;\; 2\leq k\leq M,\;\;\;\mbox{and}\;\;\; P_2\nabla h_M=0.
\ee  %
The functions $h_1, h_M$ are easy to find from the form of the Poisson
matrices:
\be\label{h1hm}  %
h_1 = \sum_{k=1}^{2M-1} J_k,\quad h_M=\prod_{k=1}^{2M-1} J_k .
\ee  %
The remaining functions, $h_2,\dots ,h_{M-1}$, are obtained by solving the
``central'' sequence of equations (\ref{ladder}).  Since $P_0$ has a Casimir
function, we need to check that the equations are compatible, in that $\nabla
h_1 P_2 \nabla h_{k-1}=0$.  We use the following result
\bl[Bi-Hamiltonian Relations] \label{biham-rels} %
With the Poisson brackets given by Definition \ref{compat-pb}, the functions
$h_1, \dots , h_M$ satisfy
$$
\{h_i,h_j\}_0=\{h_i,h_{j-1}\}_2 \quad\mbox{and}\quad
\{h_i,h_j\}_2=\{h_{i+1},h_j\}_0 .
$$
\el  %

\medskip\noindent {\bf Proof:} The ladder relations (\ref{ladder}) imply
$$
\{h_i,h_j\}_0 = \nabla h_i P_0 \nabla h_j = \nabla h_i P_2 \nabla h_{j-1} =
\{h_i,h_{j-1}\}_2,
$$
and
\bea  %
\{h_i,h_j\}_2 = \nabla h_i P_2 \nabla h_j &=& -\nabla h_j P_2 \nabla h_i \nn\\
    &=& -\nabla h_j P_0 \nabla h_{i+1} =\nabla h_{i+1} P_0 \nabla h_j =
\{h_{i+1},h_j\}_0.  \nn
\eea  %

\bl[Compatibility of Equations (\ref{ladder})] \label{compat-lad} %
Equations (\ref{ladder}) are compatible.
\el  %

\medskip\noindent {\bf Proof:} The compatibility condition $\nabla
h_1 P_2 \nabla h_{k-1}=0$ is just $\{h_1,h_{k-1}\}_2=0$.  The first equation is
just
$$
\{h_1,h_1\}_2=0,
$$
which is obviously satisfied, so it is possible to solve for $h_2$.  Now
suppose we have functions $h_1,\dots ,h_{k-1}$.  The compatibility condition is
$$
\{h_1,h_{k-1}\}_2=\{h_2,h_{k-1}\}_0=\{h_2,h_{k-2}\}_2= \cdots
=\{h_s,h_s\}_\ell=0,
$$
for some $s,\ell$.

\bigskip

To solve Equations (\ref{ladder}) write the equations in terms of the
coordinates $(J_0,\dots ,J_{2M-3}, z_1)$ (with $z_1=h_1$), after which $P_0$
has a complete row (and column) of zeros, with the non-zero part being
invertible. The above compatibility means that the final entry in the column
vector $P_2\nabla h_{k-1}$ is zero. These calculations are straightforward and
give rise to a sequence of functions of $(J_0,\dots ,J_{2M-3}, z_1)$.  These
are only defined up to an additive function of $z_1$, which can be discarded.
Replacing $z_1$ by $h_1$, we obtain the desired functions of $(J_0,\dots
,J_{2M-3}, J_{2M-2})$. We then have the following theorem of \cite{78-4}:
\bt[Complete Integrability]\label{involut}   %
The functions $h_1, \dots , h_M$ are in involution with respect to both of the
above Poisson brackets
$$
\{h_i,h_j\}_0 = \{h_i,h_j\}_2 = 0,\quad\mbox{and hence}\quad \{h_i,h_j\}_P = 0.
$$
It then follows from Liouville's theorem that the functions $h_1,\dots ,h_M$
define a completely integrable Hamiltonian system.
\et  %

\noindent {\bf Proof:}  Without loss of generality, choose $i<j$.  Then
$$
\{h_i,h_j\}_0=\{h_i,h_{j-1}\}_2= \{h_{i+1},h_{j-1}\}_0= \cdots =
\{h_k,h_k\}_\ell=0,
$$
for some $k,\ell$.  Similarly
$$
\{h_i,h_j\}_2=\{h_{i+1},h_j\}_0= \{h_{i+1},h_{j-1}\}_2= \cdots =
\{h_k,h_k\}_\ell=0,
$$
for some $k,\ell$.

\subsubsection*{The Casimir Function}  %

Formula (\ref{biham}) just states that the function
\be  \label{casimir}  %
{\cal C} = h_M - h_{M-1} + h_{M-2}-\dots +(-1)^{M+1} h_1
\ee  %
is the Casimir of the Poisson matrix $P$, so (with respect to $\{\;\;
,\;\;\}_P$) commutes with each $J_i$ and hence with {\em all} functions of
$J_i$ (not just with $h_1,\dots ,h_M$).

\bex[The Case $N=4$]  {\em   %
Here we have $3$ basic functions $J_0, J_1, J_2$.  With, $M=2$, we have $h_1$
and $h_2$, given by (\ref{h1hm}).
}\eex  %

\bex[The Case $N=6$]  {\em   %
Here we have $5$ basic functions $J_0, \dots , J_4$.  With, $M=3$ we have $h_1$
and $h_3$, given by (\ref{h1hm}), and
$$
h_2 = J_0J_1J_2+J_1J_2J_3+J_2J_3J_4+J_3J_4J_0+J_4J_0J_1.
$$
}\eex  %

\bex[The Case $N=8$]  {\em   %
Here we have $7$ basic functions $J_0, \dots , J_6$.  With $M=4$, we have $h_1$
and $h_4$, given by (\ref{h1hm}), and
$$
h_2 = \sum_{i=0}^{6} J_{i}J_{i+1}(J_{i+2}+J_{i+4}),\qquad
  h_3= \sum_{i=0}^{6} J_iJ_{i+1}J_{i+2}J_{i+3}J_{i+4},
$$
the indices here being taken modulo $6$.
}\eex  %

\br  %
It can be seen from the list following Theorem \ref{linear-thm}, that $S_4,
S_6$ (replacing $c_i$ by $J_i$) are just $\cal C$ of the above examples. We can
use the linear difference equation (\ref{lineq}) to {\bf define} $S_N$,
repeatedly using the formula $x_nx_{n+N}=x_{n+1}x_{n+N-1}+1$ to rewrite this as
a function of $x_0,\dots ,x_{N-1}$.
\er   %

\bigskip\noindent {\bf Conjecture:} The Casimir function for general $N$ can
also be written as
\be  \label{casimir2}  %
{\cal C} =  \frac{x_0+x_{2(N-1)}}{x_{N-1}} \quad\mbox{written in terms of}\;\;
x_0,\dots ,x_{N-1} .
\ee   %

\section{The Maps in Canonical Coordinates}\label{canonical}

The Poisson bracket (\ref{log-can}), with the $P_{ij}$ being given by Lemma
\ref{lemma-pij}, naturally separates the odd and even numbered variables, from
which we construct respectively canonical variables $p_i$ and $q_i$ as follows:
\bea    %
&& q_i=\log(x_{2(i-1)}), \;\; i=1,\dots ,M , \label{can-qi} \\
&& p_1=\frac{1}{2}\log(x_1x_{N-1}), \quad
p_i=\frac{1}{2}\log\left(\frac{x_{2i-1}}{x_{2i-3}}\right),
    \;\; i=2,\dots ,M ,   \label{can-pi}
\eea   %
where $N=2M$.  Defining
$$
\pi_r = \sum_{i=1}^r p_i - \sum_{i=r+1}^M p_i, \;\;\; 0\leq r\leq M-1,\;\;\;
   \pi_M =\sum_{i=1}^M p_i ,
$$
(so $\pi_i=\log(x_{2i-1})$) the inverse of this transformation is written
\be   \label{can-var2}  %
 x_{2r}=e^{q_{r+1}},\quad x_{2r+1}=e^{\pi_{r+1}},\;\;\; 0\leq r\leq M-1,
\ee   %
and the functions $J_k$ take the form
\bea   %
&&  J_{2r}=e^{-\pi_{r+1}} (e^{q_{r+1}}+e^{q_{r+2}}),
      \;\; 0\leq r\leq M-2 ,   \nn\\[2mm]
&&   J_{2r+1}=e^{-q_{r+2}} (e^{\pi_{r+1}}+e^{\pi_{r+2}}),
    \;\; 0\leq r\leq M-2,   \label{jq} \\[2mm]
&&      J_{2M-2} = e^{-q_1-\pi_M} (e^{q_1+q_M}+e^{\pi_1+\pi_M}+1).\nn
\eea  %
The map $\varphi$ (see (\ref{mapN})) is canonical, now having the form
\be  \label{phican}  %
\begin{array}{l}
\tilde q_r = \pi_r ,\; 1\leq r\leq M \\[3mm]
     \tilde p_1 =\frac{1}{2}(q_2-q_1+Log(1+e^{2p_1})) ,\quad
         \tilde p_M  = \frac{1}{2} (-q_1-q_M +\log(1+e^{2p_1})) , \\[3mm]
\tilde p_r = \frac{1}{2} (q_{r+1}-q_r) ,\;\; 2\leq r\leq M-1.
\end{array}
\ee  %
The variables $\pi_r$, transform as
\be  \label{pitilde}  %
\tilde \pi_r = q_{r+1},\;\; 1\leq r\leq M-1,\;\;\;
     \tilde \pi_M =  -q_1 +\log(1+e^{(\pi_1+\pi_M)}) .
\ee  %
The functions (\ref{jq}) inherit the cyclic behaviour (\ref{jophi}) under this
map.

Now consider the function
\bea    %
{\cal C} &=& \sum_{i=1}^{M-1} e^{-\pi_i} (e^{-q_i}+e^{-q_{i+1}}) +
    e^{-\pi_M}(e^{q_1}+e^{-q_M})+e^{\pi_M-q_1} \label{casq}  \\[3mm]
    &=&   \sum_{i=1}^{M-1} e^{-q_{i+1}} (e^{-\pi_i}+e^{-\pi_{i+1}}) +
    e^{-q_1}(e^{-\pi_1}+e^{\pi_M})+e^{q_1-\pi_M}.  \nn
\eea  %
The second line is just a re-ordering of the first, but useful.

\bl[Symmetry under the map (\ref{phican})]  %
Under the map (\ref{phican}), the function $\cal C$ is invariant: $\tilde {\cal
C}={\cal C}$.
\el  %

\smallskip\noindent {\bf Proof}

\smallskip\noindent  Using (\ref{pitilde}) it is easy to show that
$$
\begin{array}{l}
e^{-\pi_i} (e^{-q_i}+e^{-q_{i+1}}) \rightarrow
    e^{-q_{i+1}} (e^{-\pi_i}+e^{-\pi_{i+1}}), \\[3mm]
        e^{\pi_M-q_1} \rightarrow e^{-q_1}(e^{-\pi_1}+e^{\pi_M}),
        \quad
    e^{-\pi_M}(e^{q_1}+e^{-q_M}) \rightarrow e^{q_1-\pi_M},
\end{array}
$$
so the first line of (\ref{casq}) transforms to the second, giving the result.

\bt[Casimir Function]  %
The function $\cal C$ is a Casimir function for the Poisson algebra of
functions $J_i$.
\et  %

\smallskip\noindent {\bf Proof}

\smallskip\noindent  First note that
$$
\left\{J_0,e^{-\pi_i} (e^{-q_i}+e^{-q_{i+1}})\right\}=
   \left\{J_0,e^{-\pi_M}(e^{q_1}+e^{-q_M})\right\}=
      \left\{J_0,e^{\pi_M-q_1}\right\}= 0 ,
$$
so $\{J_0,{\cal C}\}=0$.  Since $\cal C$ is an invariant function under the map
(\ref{phican}), this implies that $\{J_r,{\cal C}\}=0$, for all $r$, giving the
result.

\br  %
We now have $3$ expressions for the Casimir function of the $J$ algebra
((\ref{casimir}), (\ref{casimir2}) and (\ref{casq})), which coincide on all
known explicit examples.  However, I have no proof that these are the same.
\er  %

\section{The B\"acklund Transformation for Liouville's Equation}\label{backlund}

Here we consider the Hamiltonian flows generated by the Casimir $\cal C$ and
the first Hamiltonian $h_1$.  Suppose these flows are respectively
parameterised by $x$ and $t$, so we have
$$
f_{x}=\{f,{\cal C}\}, \quad  f_{t}=\{f,h_1\},\quad\mbox{for any function}\;\;\;
   f(q_1,\dots, p_M).
$$
Since these Hamiltonians Poisson commute, their respective flows commute, so
can be considered as coordinate curves on the level surface given by ${\cal
C}=c_1, h_1=c_2$. Consider the second order partial derivative
$$
q_{ixt}=\{\{q_i,{\cal C}\},h_1\}=\{\{q_i,h_1\},{\cal C}\}.
$$
To calculate this in general, we need the formula
$$
\{q_i,\pi_j\}=\left\{\begin{array}{rl}
                             1 &\mbox{if}\;\; i\leq j , \\
                             -1 &\mbox{if}\;\; i\geq j+1 .
                             \end{array}  \right.
$$
First consider $q_1$.  From the definitions (\ref{jq}), we have
\be  \label{q1j}  %
\begin{array}{l}
\{q_1,J_{2r}\}=-J_{2r},\quad  \{q_1,J_{2r+1}\}=J_{2r+1},
\;\;\; 0\leq r\leq M-2\\[3mm]
 \{q_1,J_{2M-2}\}=-J_{2M-2}+2 e^{\pi_1-q_1} .
 \end{array}
\ee   %
Since $\cal C$ commutes with all $J_k$,
$$
\{\{q_1,h_1\},{\cal C}\}=2\{e^{\pi_1-q_1},{\cal C}\} .
$$
We have
\bea   %
&&  \{e^{\pi_1-q_1},e^{-\pi_1} (e^{-q_1}+e^{-q_{2}})\} = 2e^{-2q_1}, \nn\\
&&  \{e^{\pi_1-q_1},e^{-\pi_i} (e^{-q_i}+e^{-q_{i+1}})\} =0,
    \;\;\mbox{for}\;\; i\neq 1,   \label{pi1-q1c}  \\
&& \{e^{\pi_1-q_1},e^{-\pi_M}(e^{q_1}+e^{-q_M})\} = 0,\quad
             \{e^{\pi_1-q_1},e^{\pi_M-q_1}\} = 0,  \nn
\eea  %
so
$$
q_{1xt}=\{\{q_1,h_1\},{\cal C}\}= 4 e^{-2q_1} .
$$
Now act with $\varphi$ on this equation (recalling the formulae (\ref{phican})
and (\ref{pitilde})) to obtain Liouville's equation for each of $q_i, \pi_i$:
\be  \label{liouville}  %
q_{ixt}= 4 e^{-2q_i},\qquad \pi_{ixt}= 4 e^{-2\pi_i},\quad i=1,\dots , M.
\ee   %
The B\"acklund transformation for this equation is well known (see
\cite{02-4}), but here we show how to construct it from our canonical
transformation (\ref{phican}) and (\ref{pitilde}).

Again, first consider $q_1$ and $\tilde q_1 = \pi_1$.  Looking at the formulae
(\ref{pi1-q1c}), we see that $q_1-\pi_1$ commutes with all but one term in the
expression (\ref{casq}) for $\cal C$.  The remaining term gives
$$
\{q_1-\pi_1,{\cal C}\}=\{q_1-\pi_1,e^{-\pi_1} (e^{-q_1}+e^{-q_{2}})\} =
-2e^{-q_1-\pi_1} .
$$
We now use (\ref{q1j}), together with their consequence under the map
\bea  %
&& \{\pi_1,J_{0}\}=J_{0}-2e^{q_1-\pi_1},\quad
\{\pi_1,J_{2r}\}=J_{2r},\;\;\mbox{for}\;\; r\neq 0,\nn\\[3mm]
&&  \{\pi_1,J_{2r+1}\}=-J_{2r+1},\quad \{\pi_1,J_{2M-2}\}=J_{2M-2},\nn
\eea  %
so
$$
\{q_1+\pi_1,h_1\}=2 (e^{\pi_1-q_1}-e^{q_1-\pi_1}).
$$
In summary, we have shown

\be  \label{bt} %
q_{1x}-\tilde q_{1x} = -2e^{-q_1-\tilde q_1},\qquad
  q_{1t}+\tilde q_{1t} = 2 (e^{\tilde q_1-q_1}-e^{q_1-\tilde q_1}),
\ee  %
which is the B\"acklund transformation for Liouville's equation
(\ref{liouville}) (for $i=1$).

Again, act with $\varphi$ on these equations to obtain
\bea   %
&& q_{ix}- \pi_{ix} = -2e^{-q_i- \pi_i},\qquad
  q_{it}+ \pi_{it} = 2 (e^{\pi_i-q_i}-e^{q_i-\pi_i}),
                     \quad i=1,\dots , M   ,  \nn\\[3mm]
&& \pi_{ix}- q_{i+1\,x} = -2e^{-\pi_i- q_{i+1}},
                     \quad i=1,\dots , M-1, \nn\\[3mm]
 && \pi_{it}+ q_{i+1\,t} = 2 (e^{q_{i+1}-\pi_i}-e^{\pi_i-q_{i+1}}),
             \quad i=1,\dots , M-1 .   \nn
\eea   %
We can act again with $\varphi$, but the calculation is more
complicated, since it now involves $\tilde \pi_M$ (see (\ref{pitilde})).  We
get a relationship involving derivatives of $3$ variables ($\pi_M, q_1$ and
$\pi_1$), but use (\ref{bt}) to eliminate derivatives of $\pi_1=\tilde
q_1$, to obtain
$$
\pi_{Mx}+q_{1x}= -2 (e^{q_1-\pi_M}-e^{\pi_M-q_1}),\qquad
   \pi_{Mt}-q_{1t}=2 e^{-\pi_M-q_1}.
$$
Notice that $x, t$ seem to have reversed their roles at this step.  However,
the next action of $\varphi$ (again requiring more complicated manipulations)
brings us full circle to the original formulae (\ref{bt}) for $q_1, \pi_1$.

\section{Conclusions}

In \cite{f09-1} a new class of quiver, with a certain periodicity property, was
introduced and partially classified.  The corresponding cluster mutation
relations give rise to iterations with the Laurent property.  An important open
question is the classification of the subclass of such iterations which define
Liouville integrable maps.  The main content of this paper is the study of one
particular family of such maps.  The question of integrability for the general
class is considered in \cite{f10-2}.

In \cite{f09-1} we noted a surprising connection between our examples of
periodic quivers and those which arise in the context of {\em quiver gauge
theories} (see \cite{05-5}).  Unfortunately, I had no space to describe this
here, but an explanation of this is also an important open question.

This brings us back, finally, to Robin Bullough's famous diagram.  Some new
boxes and connections are needed to incorporate the subject of this paper, but
that is the nature of this diagram, which will grow indefinitely and become
more and more complex as new discoveries are made.


\begin{thebibliography}{14}
\providecommand{\natexlab}[1]{#1}
\providecommand{\url}[1]{\texttt{#1}}
\expandafter\ifx\csname urlstyle\endcsname\relax
  \providecommand{\doi}[1]{doi: #1}\else
  \providecommand{\doi}{doi: \begingroup \urlstyle{rm}\Url}\fi

\bibitem[Bellon and Viallet(1999)]{99-12}
M.~Bellon and C-M. Viallet.
\newblock Algebraic entropy.
\newblock \emph{Comm.Math.Phys.}, 204:\penalty0 425--37, 1999.

\bibitem[Fomin and Zelevinsky(2002{\natexlab{a}})]{02-2}
S.~Fomin and A.~Zelevinsky.
\newblock The {Laurent} phenomenon.
\newblock \emph{Advances in Applied Mathematics}, 28:\penalty0 119--144,
  2002{\natexlab{a}}.

\bibitem[Fomin and Zelevinsky(2002{\natexlab{b}})]{02-3}
S.~Fomin and A.~Zelevinsky.
\newblock Cluster algebras {I}: Foundations.
\newblock \emph{J. Amer Math Soc}, 15:\penalty0 497--529, 2002{\natexlab{b}}.

\bibitem[Fordy and Hone(2010)]{f10-2}
A.P. Fordy and A.N.W. Hone.
\newblock Integrable maps and {Poisson} algebras derived from cluster algebras.
\newblock 2010.
\newblock In preparation.

\bibitem[Fordy and Marsh(2009)]{f09-1}
A.P. Fordy and R.J. Marsh.
\newblock Cluster mutation-periodic quivers and associated laurent sequences.
\newblock 2009.
\newblock Preprint arXiv:0904.0200v3 [math.CO]. To appear in the Journal of
  Algebraic Combinatorics.

\bibitem[Gale(1991)]{91-17}
D.~Gale.
\newblock The strange and surprising saga of the {Somos} sequences.
\newblock \emph{Math. Intelligencer}, 13:\penalty0 40--2, 1991.

\bibitem[Gekhtman et~al.(2003)Gekhtman, Shapiro, and Vainshtein]{03-5}
M~Gekhtman, M~Shapiro, and A~Vainshtein.
\newblock Cluster algebras and {Poisson} geometry.
\newblock \emph{Moscow Math.J}, 3:\penalty0 899--934, 2003.

\bibitem[Hanany et~al.(2005)Hanany, Kazakopoulos, and Wecht]{05-5}
A.~Hanany, P.~Kazakopoulos, and B.~Wecht.
\newblock A new infinite class of quiver gauge theories.
\newblock \emph{J. High Energy Phys.}, 08:\penalty0 054, 2005.

\bibitem[Hone(2007)]{07-2}
A.N.W. Hone.
\newblock Laurent polynomials and superintegrable maps.
\newblock \emph{SIGMA}, 3:\penalty0 022, 18 pages, 2007.

\bibitem[Magri(1978)]{78-4}
F.~Magri.
\newblock A simple model of the integrable {Hamiltonian} equation.
\newblock \emph{J.Math.Phys}, 19:\penalty0 1156--1162, 1978.

\bibitem[Quispel et~al.(1988)Quispel, Roberts, and Thompson]{88-5}
G.R.W. Quispel, J.A.G. Roberts, and C.J. Thompson.
\newblock Integrable mappings and soliton equations.
\newblock \emph{Phys. Letts. A.}, 126:\penalty0 419--421, 1988.

\bibitem[Rogers and Schief(2002)]{02-4}
C.~Rogers and W.K. Schief.
\newblock \emph{{B\"acklund and Darboux Transformations: Geometry and Modern
  Applications in Soliton Theory}}.
\newblock CUP, Cambridge, 2002.

\bibitem[Sloane(2009)]{09-1}
N.J.A. Sloane.
\newblock {The On-Line Encyclopedia of Integer Sequences}.
\newblock \emph{www.research.att.com/$\sim$njas/sequences}, 2009.

\bibitem[Veselov(1991)]{91-4}
A.P. Veselov.
\newblock Integrable maps.
\newblock \emph{Russ. Math. Surveys}, 46, N5:\penalty0 1--51, 1991.

\end{thebibliography}

\end{document}